\documentclass[sigconf]{acmart}
\AtBeginDocument{%
  }

\copyrightyear{2024}
\acmYear{2024}
\setcopyright{acmlicensed}
\acmConference[x 'XX]{x}{July 03--05,
  2024}{Glasgow, UK}
\acmDOI{XXXXXXX.XXXXXXX}

\usepackage{dsfont}
\usepackage[para]{footmisc}
\usepackage[show]{chato-notes}
\usepackage{enumitem}
\usepackage{hyperref}
\newcommand{\pageenlarge}[1]{\enlargethispage{#1\baselineskip}}

\newcommand{\zy}[1]{\textcolor{black}{#1}}
\newcommand{\zm}[1]{\textcolor{black}{#1}}
\newcommand{\io}[1]{\textcolor{black}{#1}}
\newcommand{\zym}[1]{\textcolor{black}{#1}}
\definecolor{darkblue}{rgb}{0.0, 0.0, 0.55}
\newcommand{\z}[1]{\textcolor{black}{#1}}
\newcommand{\zz}[1]{\textcolor{black}{#1}}
\begin{document}



\title{KERAG\_R: Knowledge-Enhanced Retrieval-Augmented Generation for Recommendation}

\author{Zeyuan Meng}
\affiliation{%
  \institution{University of Glasgow}
  \city{Glasgow}
  \country{UK}}
  \email{z.meng.2@research.gla.ac.uk}

\author{Zixuan Yi}
\affiliation{%
  \institution{University of Glasgow}
  \city{Glasgow}
  \country{UK}}
  \email{z.yi.1@research.gla.ac.uk}

\author{Iadh Ounis}
\affiliation{%
  \institution{University of Glasgow}
  \city{Glasgow}
  \country{UK}}
  \email{iadh.ounis@glasgow.gla.ac.uk}

\renewcommand{\shortauthors}{Meng et al.}

\begin{abstract}
Large Language Models (LLMs) have notably enhanced performance across various tasks, including in recommender systems, due to their strong capabilities for contextual learning and generalisation. 
Existing LLM-based recommendation approaches typically formulate the recommendation task using specialised prompts designed to leverage their contextual abilities, and aligning their outputs closely with human preferences to yield an improved recommendation performance. 
However, the use of LLMs for recommendation tasks is limited by the absence of domain-specific knowledge. 
This lack of relevant relational knowledge about the items to be recommended in the LLM’s pre-training corpus can lead to inaccuracies or hallucinations, resulting in incorrect or misleading recommendations. 
Moreover, directly using information from the knowledge graph introduces redundant and noisy information, which can affect the LLM’s reasoning process or exceed its input context length, thereby reducing the performance of LLM-based recommendations. 
To address the lack of domain-specific knowledge, we propose a novel model called Knowledge-Enhanced Retrieval-Augmented Generation for Recommendation (KERAG\_R).
Specifically, we leverage a 
graph retrieval-augmented generation (GraphRAG)
component to integrate additional information from a knowledge graph (KG) into instructions, enabling the LLM to collaboratively exploit recommendation signals from both text-based user interactions and the knowledge graph to better estimate the users’ preferences in a recommendation context. 
In particular, we perform graph RAG by pre-training a graph attention network (GAT) to select the most relevant triple for the target users for the used LLM, thereby enhancing the LLM while reducing redundant and noisy information. 
Moreover, we leverage a knowledge-enhanced instruction tuning approach to incorporate relational knowledge during the LLM’s tuning stage, thereby enhancing the adaptation of the used LLM in recommender systems.
Our extensive experiments on three public datasets show that our proposed KERAG\_R model significantly outperforms ten existing state-of-the-art recommendation methods. 
In particular, our KERAG\_R model outperforms the best baseline, namely RecRanker, an LLM-based recommendation model by up to 14.89\% on the Amazon-Book dataset. 
We also find that using graph RAG to retrieve the most relevant KG information is more effective than using additional KG triples, and that relational KG triple representations outperform natural KG sentence representations in the prompts.
\pageenlarge{4}
\end{abstract}



\maketitle

\pageenlarge{4}
\section{Introduction}\label{s1}
Large Language Models (LLMs)~\cite{touvron2023llama,zhao2023survey,achiam2023gpt} have demonstrated remarkable capabilities in text understanding~\cite{luo2024large}, generation~\cite{li2023large}, and reasoning~\cite{amirizaniani2024can}.
\io{Leveraging} their strong generalisation ability, LLMs have been actively integrated into various domains, including recommender systems, to enhance personalisation and user experience~\cite{zhao2024recommender,shang2024personalized}.
Recommender systems
\zz{~\cite{wang2020time,liu2019user,gharibshah2021user,yi2022multi,yi2023contrastive}}
\io{is a} widely used application, \io{which aims} to recommend potential items to the target users \io{in} various online services\zz{.}
\zz{In particular,}
LLMs have been incorporated \io{into} various recommendation tasks, including \io{collaborative filtering} and sequential recommendation~\cite{wei2024llmrec,zhang2023recommendation, yang2023palr}.
\zz{Early LLM-based recommendation models encoded user features from historical interactions~\cite{sileo2022zero,zhang2021language}, 
\zz{but often struggled}
to process the full interaction history 
within a single prompt due to 
\zz{the limited input token length of LLMs}
~\cite{geng2022recommendation,shin2021one4all}.}
\zym{To address this limitation, later LLM-based recommendation models proposed filtering 
\zz{the most relevant}
textual information of items and users input to LLMs, 
thereby improving \io{the} recommendation performance~\cite{cui2022m6,wang2023zero}.}
\zym{Despite these improvements,}
\zym{these \io{enhanced} approaches still use}
LLMs directly without \zym{specific} fine-tuning 
\zym{for the recommendation tasks.}
\zym{This general application of LLMs without adaptation \io{to the task at hand} can misalign the models with their \io{actual} recommendation task, thereby limiting the performance of these models~\cite{wang2023zero,sileo2022zero}.}
\zym{\io{Instead, more} recent studies} 
~\cite{zhang2023recommendation,luo2023recranker,yi2025enhancing,yi2025multi,yi2023large} have proposed \zym{using natural language instruction adaptation to better align the LLMs to specific recommendation tasks.}
\zym{RecRanker~\cite{luo2023recranker} is a typical example, which uses adaptive user sampling to construct representative prompts for instruction tuning to align \io{the} Llama-2 model~\zym{\cite{touvron2023llama}} with the top-$k$ recommendation task \io{at hand}.}
\zym{\io{Nevertheless}, \io{the} existing LLM-based recommendation models still predominantly depend on \io{the content of the} user interactions and \io{the} LLM's pre-training corpus, \io{and might suffer from} the absence of domain-specific knowledge \io{to support the recommendation task at hand}.}
Although LLMs \io{have} excellent context \io{representation} abilities, the lack of relevant 
\zym{relational}
knowledge about the items to be recommended in \io{the} LLM’s pre-training corpus can cause \io{these LLMs} to “hallucinate” and thus produce incorrect or misleading recommendations.
\zym{\z{We argue that the adequate integration of} external knowledge sources into the LLM-based recommendation models is important to enhance the \io{LLMs'} ability to learn specific domain knowledge in a recommendation scenario.}

\zym{Recently, Retrieval-augmented Generation (RAG) has been proposed to alleviate the problem of LLM’s lack of domain-specific knowledge by querying external 
sources
and incorporating relevant factual knowledge into \io{the} generated responses~\cite{peng2024graph,hu2024grag}.}
\zym{Some works~\cite{borgeaud2022improving,guu2020retrieval} \io{used} RAG to capture textual content and information in a text corpus to complement \io{the} LLMs 
\zym{with domain-specific knowledge.}}
\zym{However, these RAG-based methods typically rely on the input textual query to 
\zym{retrieve }
the textual information, ignoring the structural relational information between text contents~\cite{lewis2020retrieval}.}
\zym{To bridge this gap,}
Graph Retrieval-Augmented Generation (GraphRAG)~\cite{edge2024local,hu2024grag} has been proposed to obtain relational information from \zym{graphs.}
\zym{Unlike conventional RAG methods, GraphRAG incorporates graph learning into RAG, \io{allowing to retrieve} graph elements -- such as nodes, triples, or subgraphs -- that contain relational knowledge relevant to a given query from a pre-constructed graph database~\cite{peng2024graph}.}
\zym{This GraphRAG method can leverage various types of graph data for retrieval and generation, including knowledge graphs~\cite{xu2024retrieval}.}
\zym{Knowledge graphs (KGs)}~\cite{guo2020survey,wang2021dskreg}, \io{which are} rich in item-related entity relational triples, provide domain-specific knowledge that, \z{we argue, could enhance} the user/item representations in 
\zz{top-$k$}
recommender systems.
\zym{Building on the GraphRAG method,}
we propose a novel recommendation model named Knowledge-Enhanced Retrieval-Augmented Generation for 
\zz{effective}
Recommendation (KERAG\_R).
Specifically, we leverage a 
\zz{GraphRAG}
component to integrate external knowledge from \io{a} KG into LLM prompts, enabling the used LLM (\io{e.g.}, Llama 3.1) to learn domain-specific knowledge tailored for a top-$k$ recommendation task, thereby improving the used LLM's ability to \io{capture} context and infer user preferences.
Moreover, to reduce the possible redundant and noisy 
\zym{relational information in the prompt}, we incorporate a graph attention network (GAT)-based KG triple selection method into the 
\z{GraphRAG component.}
This selection method selects the knowledge graph triples \io{that are} most relevant to \io{the} user interactions, reducing noise and redundant information, thereby enhancing the recommendation performance of the LLM-based model.
\z{Additionally, we propose a knowledge-enhanced
instruction tuning approach to} 
\zz{further enable the LLM to 
\zz{to incorporate structured knowledge about item relations}
from the KG during the tuning stage.} \looseness -1
\pageenlarge{3}
Overall, the contributions of our work are summarised as follows:
(1) \zym{Our \io{new} KERAG\_R model leverages a GraphRAG component to integrate external knowledge from \io{a} KG into LLM prompts, thereby enhancing LLM's reasoning in the top$-k$ recommendation task.}
To reduce the introduction of redundant and noisy information \io{from} the KG, we use a KG triple selection method to select the most relevant triple for the target users.
\z{To enhance the adaptation of LLMs in recommender systems,} 
\zz{we incorporate relational knowledge through a knowledge-enhanced instruction tuning approach during the LLM’s tuning stage;}
(2) \io{We} conduct extensive experiments on three public datasets to evaluate our proposed KERAG\_R model.
We show that our KERAG\_R model significantly outperforms 
\zz{ten}
existing state-of-the-art recommendation models, in particular outperforming \zym{the strongest baseline}, RecRanker, across \io{the} three used public datasets; 
(3)
\io{We conduct an} ablation study \io{that confirms} the effectiveness of KERAG\_R's 
\zym{GraphRAG component \io{as well as our} KG triple selection method;}
(4)
Moreover, we \io{show} that retrieving \z{the} most relevant triple for each interaction of each user performs better than using additional \io{KG triples}, 
\zym{and that relational KG triple representations outperform natural KG sentence representations in the prompts.}\looseness -1

\section{Related Work}\label{s2}
In this section, we position our work and introduce two related works that are relevant to our work, namely 
\zym{LLM-based recommendation and retrieval-augmented generation.}

\subsection{LLM-based Recommendation}\label{s2.1} 
Recently, large language models (LLMs) have demonstrated strong reasoning capabilities and have found applications across various domains, including recommender systems~\cite{geng2022recommendation,sun2023chatgpt,yi2024directional,yi2023graph,meng2024knowledge,yi2024unified,meng2024knowledge,yi2025multi}.
In the context of recommendation models, LLM applications typically employ either zero-shot or fine-tuned approaches.
Zero-shot methods leverage the inherent capabilities of LLMs to handle recommendation tasks without any specific model training.
These methods typically \io{use} prompts derived from user interaction data to guide the LLMs \io{in predicting the} users' preferences.
For example, Sileo et al.~\cite{sileo2022zero} extracted user/item side information -- such as \z{movie titles} -- from textual descriptions to create prompts, and then used GPT-2\cite{radford2019language} to predict potential \io{items} for the target users.
\zz{However,} 
zero-shot methods may face challenges due to a fundamental misalignment between the LLMs’ general capabilities and the specific demands of \io{specific} recommendation tasks, often requiring additional model adjustments. 
\zz{On the other hand,}
fine-tuned methods involve explicitly adapting \io{the} LLMs to \io{specific} recommendation tasks, \io{and were} generally \io{shown to be} more effective than zero-shot approaches~\cite{kang2023llms,cui2022m6}.
\z{For instance, Zhang et al.~\cite{zhang2023recommendation} developed an instruction 
\zz{template}
designed to 
\zz{organise}
task inputs for \io{the} LLMs during the fine-tuning phase -- typically \io{this} includes a task definition, \io{an} input context, and \io{an} output specification -- effectively guiding the LLM to 
\zz{perform}
specific recommendation tasks.}
RecRanker~\cite{luo2023recranker} 
\io{further enhanced} this instruction format by incorporating adaptive user sampling to generate more effective prompts during the instruction tuning stage, \io{in particular} adapting the LLM for top-$k$ 
ranking tasks,
\zz{such as}
\io{a} listwise \io{ranking} task~\cite{xia2008listwise}.
However, these fine-tuned models still predominantly rely on user interaction data and the broad knowledge base of the LLM’s pre-training corpus, which may lack the domain-specific information necessary for accurate recommendations. This \io{inherent limitation can also} lead to errors \io{such as the} well-known “hallucination” problem, where LLMs generate incorrect recommendations for the target users~\cite{ji-etal-2023-rho}.
To address the absence of domain-specific knowledge, we propose a knowledge-enhanced instruction tuning approach 
that incorporates relational knowledge (i.e., item-entity triples) during the LLM's tuning stage, thereby \io{supporting the} LLMs with additional \io{knowledge to tackle specific recommendation tasks}.  
To the best of our knowledge, we are the first 
\z{to incorporate relational knowledge in instruction tuning for \io{effective} LLM-based recommendation models.}

\pageenlarge{3}
\subsection{\zym{Retrieval-Augmented Generation}}\label{s2.2}
Recently, Retrieval-Augmented Generation (RAG) has demonstrated its ability to 
\z{capture}
large corpora of text from external sources, incorporating domain-specific knowledge into the responses generated by the \io{large} language model, thereby 
\z{enabling LLMs to produce more cohesive and contextually relevant responses~\cite{peng2024graph,fan2024survey,gao2023retrieval,hu2024rag}.}
\z{Indeed,} RAG aims to capture textual content and information, which are contextually relevant to a given textual query~\cite{hu2024grag,guu2020retrieval,wu2024coral}.
\zz{For example, REALM~\cite{guu2020retrieval} employed a RAG method to retrieve relevant documents for each input query 
\zz{from external sources,}
thereby improving their performance on the Open-QA task.}
However, these RAG methods primarily rely on retrieving textual content based on 
\zz{query-text matching}
overlooking
the 
\zz{underlying}
structural and relational information between \io{textual} contents.
This oversight can result in the generation of inaccurate responses by \io{the} LLMs~\cite{lewis2020retrieval,liu2024lost}.
To address this limitation, Graph Retrieval-Augmented Generation (GraphRAG)~\cite{edge2024local,hu2024grag} has been \io{recently} proposed to capture \zym{structural relational information} \z{between \io{textual} contents within \io{the} LLM's prompts.}
Unlike conventional RAG, GraphRAG considers the interconnections among the key entities within \io{the} textual contents, thereby enhancing the quality of the generated responses by leveraging \io{such} structural relationships~\cite{peng2024graph,mavromatis2024gnn}.
For example, G-Retriever~\cite{he2024g} \io{introduced} the first RAG method for textual graphs, which \io{allows the LLM to be} fine-tuned to enhance graph understanding
in \z{the} OpenQA task.
Therefore, we propose the use of the GraphRAG method with knowledge graphs to integrate 
\zz{LLMs with}
domain-specific knowledge in a top-$k$ recommendation scenario. 
To the best of our knowledge, we are the first to apply a GraphRAG method
to LLM-based 
\zz{top-$k$ recommendation task.}
In addition, in order to reduce noise and redundant information \io{that could be introduced by} the used KG, we leverage a 
\zz{pre-trained}
graph attention network (GAT) 
model to perform a KG triples selection \io{method}, \io{which allows to} obtain the most relevant triples for \io{each of} a given user interacted \io{items}. 

\section{Methodology}\label{s3}
In this section, we start by introducing some preliminary background information (Section\zz{~\ref{s3.1}}) as well as the notations we will
be using in the remainder of the paper.
Next,
in Section\zz{~\ref{s3.2}}, we describe our proposed KERAG\_R model and its 
\zym{model}
architecture (as illustrated in Figure~\ref{fig1}). 
\zz{Sections~\ref{s3.3} and~\ref{s3.4} present our proposed GraphRAG method and the prompt construction approach 
for the top-$k$ recommendation task.}
\zz{Then,}
Section~\ref{s3.5} presents our 
\z{knowledge-enhanced}
instruction tuning 
\zm{method for top-$k$ recommendation.}
\zz{Finally, we analyse the efficiency of our model in Section~\ref{s3.6}.}\looseness -1 

\pageenlarge{3}
\subsection{\zy{Preliminaries}}\label{s3.1}
\looseness -1 A vanilla instruction tuning process in recommendation systems typically involves three key steps: 
\io{(i)} instruction prompt construction, 
\zz{(ii)}
\zz{instruction tuning the LLM}
and \io{(iii)} top-$k$ ranking.

\noindent \textbf{Instruction prompt construction:}
An instruction prompt for the top-$k$ recommendation task consists of user-item interactions, candidate items for the target user, \io{the} task description and \io{the} output specification. 
We define ${\mathcal U}$ and ${\mathcal I}$ to represent the users set and the items set, respectively: ${\mathcal U}=\{u_{1},u_{2},\ldots,u_{M}\}$ and ${\mathcal I}=\{i_{1},i_{ 2},\ldots,i_{N}\}$.
We also define the user-item \io{interactions} matrix as $\mathbf{Y}\in\mathbb{R}^{M\times N}$, where $M$ and $N$ represent the number of users and items, respectively.
\io{In} an LLM-based approach, we \io{also} need to interpret the recommendation task of user $u$ into natural language using prompts to match the input requirements of the \io{used} LLM.
\zz{Specifically, we obtain each sampled user's likes and dislikes from their historical interactions based on high and low ratings, respectively. }
\zz{To construct the candidate item list $\mathcal{S}_{u} \subset I$ for each user $u \subset\mathcal U$, we 
\zz{follow RecRanker~\cite{luo2023recranker}}
for training and inference: during training, we combine liked and disliked items with non-interacted items generated via negative sampling~\cite{rendle2012bpr,yang2020mixed}; during inference, we adopt a traditional recommender, LightGCN, to retrieve the candidate set.}
Then, we 
\zz{express}
the 
\zz{user-item interactions} 
information,
\z{along with the corresponding candidate item list,}
in natural language \zym{as the} task description. 
\zym{Finally, we constrain the LLM's output with an output specification within the instruction prompt}
to ensure the generation of the potential top-$k$ items for the target user.\looseness -1

\noindent \textbf{\zz{Instruction tuning the LLM}:}
After constructing the instruction prompts, we use $Prompt_u$ as the input query 
\zz{to perform instruction tuning, effectively fine-tuning}
the used LLM (\io{e.g.,} Llama 3) for a listwise ranking task with the cross-entropy loss\zym{~\cite{zhang2018generalized}}. 
\z{This fine-tuning process updates the model's parameters from the pre-trained state $\theta$ to the fine-tuned state $\theta^\prime$, adapting the LLM to a top-$k$ recommendation scenario.}\looseness -1

\noindent \textbf{Top-$k$ ranking:}
The objective of the top-$k$ recommendation task is to estimate the users' preferences through an LLM-based recommendation model $f_{\theta}$ (e.g., Llama-3), which recommends the top-$k$ items for the target user $u$.
After \io{the} instruction tuning 
\io{of} the LLM model, we obtain an updated LLM, $f_{\theta^\prime}$, specifically trained 
\io{on a} listwise ranking task, 
\zym{which has been shown to be a more effective method than using pairwise and pointwise ranking approaches~\cite{luo2023recranker}.}
We then input a prompt \zm{$Prompt_u$} to this updated LLM to generate a list of top-$k$ items $\mathcal{S}_{u}^{\prime}$ for each user $u\in\mathcal{U}$, based on the candidate item list $\mathcal{S}_{u}$.\looseness -1
\pageenlarge{3}
\subsection{Model Overview}\label{s3.2}
\begin{figure*}[tb]
\centering
\includegraphics[width=1.0\linewidth]{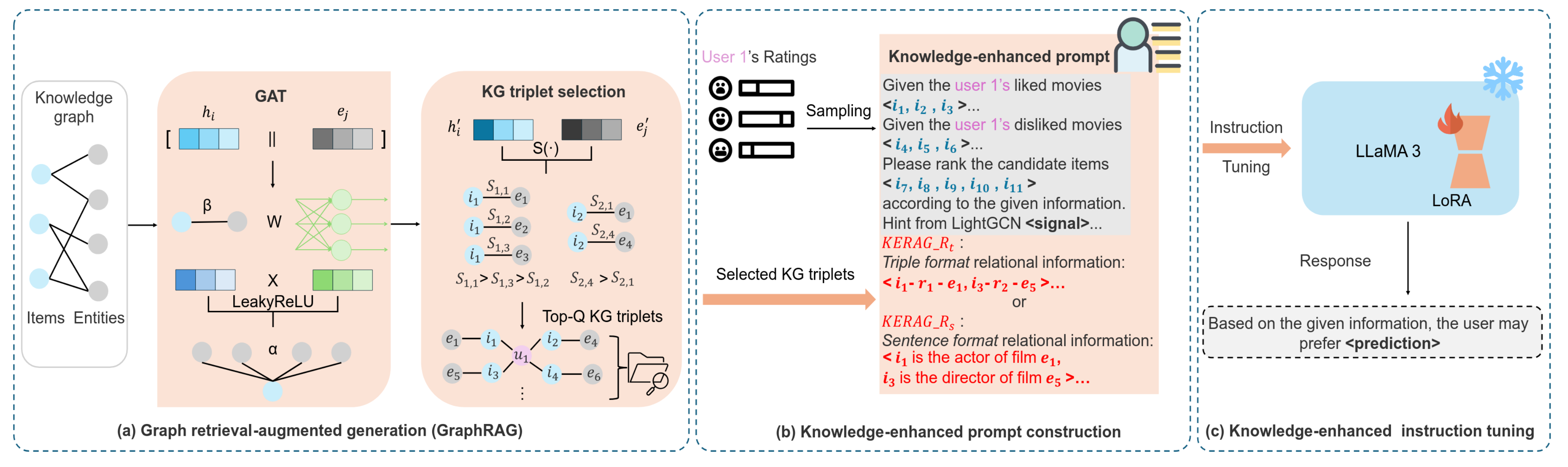}
\vspace{-6mm}
\caption{The architecture of our proposed KERAG\_R model with three parts: (a) Graph retrieval-augmented generation (GraphRAG), (b) Knowledge-enhanced prompt construction and (c) Knowledge-enhanced instruction tuning. } 
\label{fig1}
\end{figure*}
Figure~\ref{fig1} illustrates the architecture of our proposed KERAG\_R model, which involves three major components: Graph Retrieval Augmented Generation (GraphRAG), knowledge-enhanced prompt construction and knowledge-enhanced instruction tuning.
These components are executed sequentially, as illustrated in Figure 1 (a), (b), and (c), respectively.

\noindent \textbf{Graph retrieval-augmented generation:} The GraphRAG component (as illustrated in Figure 1 (a)) first retrieves the top-$Q$ 
relevant entities for each item to construct the instruction prompt (as illustrated in Figure 1 (b)).
Specifically, we use a Graph Attention Network (GAT) to pre-train the item/entity embeddings of the KG.
\zz{Using these item/entity embeddings, we compute dot product similarity scores to rank candidate entities and select the most relevant triples for each of the user's interacted items.}

\noindent \textbf{Knowledge-enhanced prompt construction:}
As shown in Figure 1 (b), after identifying the most relevant triples for each \io{of the} user's interacted items \io{using} the GraphRAG component, we incorporate these selected triples into the instruction prompt, as well as the user’s like and dislike item lists, and a primary ranking list from a recommender (i.e., LightGCN).

\noindent \textbf{Knowledge-enhanced instruction tuning:}
We then use the constructed instruction prompt, 
which incorporates item-entity triples from the GraphRAG component as relational knowledge,
to fine-tune the used LLM for the top-$k$ recommendation task, as illustrated in Figure 1 (c).
Specifically, we use Llama-3.1-8B-Instruct~\cite{dubey2024llama} as 
\zz{the}
\zz{backbone model}
\io{because of} its effectiveness in various tasks \zym{~\cite{fang2024llama,luo2024badam}}. For the fine-tuning process, we employ the parameter-efficient tuning method, LoRA~\cite{hu2021lora} to optimise Llama-3 
for this proposed knowledge-enhanced instruction tuning.
Once fine-tuned with the selected triples, we obtain an instruction-tuned LLM that generates a listwise set of items for the target user.\looseness -1

\subsection{GraphRAG for LLM-based Recommendation}\label{s3.3}
\zym{As discussed in Section\zz{~\ref{s2.2}}, we aim to retrieve the most relevant KG triples for each item and integrate them into the LLM to mitigate the absence of domain-specific knowledge.}
To achieve this, we begin by pre-training \io{the} item/entity embeddings using a GAT model for later retrieval. 
Subsequently, we retrieve the 
\z{top-$Q$}
relevant 
\z{triples}
for each item by comparing the \z{dot product} similarities 
between the updated \z{items}  and the \z{entities} embeddings \z{through the KG triple selection method}. 

\subsubsection{GAT Pre-training for Triple Retrieval}~\label{s3.3.1}
As \io{described} in Section 3.2 
, we aim to leverage a GAT\zz{~\cite{velivckovic2017graph}} model to retrieve the 
\z{top-$Q$}
relevant triples for the target user's interactions.
Initially, we input triple data from the knowledge graph ${\mathcal G}_{k}=\{(h,r,t)\}$, 
\zz{where $h$, $r$, and $t$ denote the head item, relation, and tail entity, respectively.}
\zz{We}
map items and entities to unique indices and initialise the embedding matrix using the Xavier initialiser.
Subsequently, we perform a graph aggregation operation by adaptively assigning weights to neighbour entities $j$ for each item $i$ through a self-attention mechanism as follows: 
\pageenlarge{1}
\begin{equation}
\alpha_{ij} = \frac{\exp(\text{LeakyReLU}(\beta^\top [W h_i \| W e_j]))}{\sum_{k \in \mathcal{N}(i)} \exp(\text{LeakyReLU}(\beta^\top [W h_i \| W e_k]))},
\end{equation}
\begin{equation}
h_i' = \sum_{j \in \mathcal{N}(i)} \alpha_{ij}  W e_j,
\end{equation}
where $h_i$ and $e_j$ represent the item and entity embeddings, $W$ is a learnable weight matrix, $k \in \mathcal{N}(i)$ represents the set of neighbours of item node $i$ and $h_i'$ is the \zym{updated item embedding. 
}
$||$ represents the vector concatenation operation, $\beta \in \mathbb{R}^{2d}$ is a learnable attention weight vector that computes the attention weights and $d$ is the embedding dimension. 
As such, this graph aggregation operation enables more enriched and context-aware \io{entity} embeddings from relevant neighbouring items.
Moreover, we use a contrastive loss to optimise the GAT pre-training process ~\cite{hadsell2006dimensionality}. Specifically, we select the triples that exist in the KG as \io{the} positive pairs $(h_i , e_k)$, while the triples that do not exist in the KG 
\z{are considered as}
negative pairs $(h_i , e_j)$:\looseness -1
\begin{equation}
\mathcal{L} = \frac{1}{N} \sum_{(i,j) \in E} \max(0, \phi(h_i , e_k) - \phi(h_i , e_j)),
\end{equation}
where $\phi(\cdot)$ denotes the inner products to determine the similarity between the positive and negative pairs, \z{and}
$E$ is the set of all positive samples.
As such, we can leverage this well-trained GAT model to generate effective item/entity embeddings for the subsequent triple selection.

\subsubsection{Knowledge Graph Triple Selection}~\label{s3.3.2}
Our next step is to select the triples from the knowledge graph (KG) that are most relevant to the item. This selection is based on the embeddings of the items and entities and the attention weights derived from the GAT.
Specifically, we calculate the dot product similarity between the item and entity embeddings, which is then refined by applying the attention weights as follows:
\begin{equation}
S_{ij} = \alpha_{ij} (h_i' \cdot e_j),
\end{equation}
\zym{Based on the computed similarity scores, we select the top-$Q$ entities for each item to obtain the final triples set:}
\begin{equation}
T_i=\mathrm{Top-Q}(S_{ij}),\quad e_j\in\mathcal{N}(i),
\end{equation}
\zy{where $T_i$ represents the 
\zym{Q} 
highest-scoring entities associated with each item.}
\zym{As such, we retrieve the top-$Q$ related triples for each item by similarity calculation,}
\zym{thereby alleviating the noise and redundant information
in KG.}

\pageenlarge{1}
\subsection{\zym{Knowledge-Enhanced} Prompt Construction}\label{s3.4}
\begin{table}[t]
\centering
\scriptsize  
\setlength{\tabcolsep}{3pt}
\caption{Illustrative examples of instruction for ranking task.}
\begin{tabular}{|l|p{7.2cm}|}
\hline
\textbf{Variant} & \textbf{Instruction} \\ \hline

\textbf{Original} & The historical interactions of a user include: \underline{\textless historical interactions\textgreater}. 
\newline How would the user rank the \underline{\textless candidate item list\textgreater}? \\ \hline

\textbf{KERAG\_R$_t$} & You are a movie recommender system. Your task is to rank a given list of candidate movies based on user preferences and return the top five recommendations.  
\newline User's Liked movies: \textbf{\underline{\textless  liked historical interactions\textgreater}}. 
\newline User's Disliked movies: \textbf{\underline{\textless disliked historical interactions\textgreater}}.
\newline Question: How would the user rank the candidate item list: \underline{\textless candidate item list\textgreater}?
\newline Hint 1: Another recommender model suggests  \textbf{\underline{\textless ranking list\textgreater}}.
\newline Hint 2: These are corresponding entities and relationships for above model’s recommendation for more context information: \textbf{\textless KG triple format information \textgreater}. 
\\ \hline

\textbf{KERAG\_R$_s$} & You are a movie recommender system. Your task is to rank a given list of candidate movies based on user preferences and return the top five recommendations.  
\newline User's Liked movies: \textbf{\underline{\textless  liked historical interactions\textgreater}}. 
\newline User's Disliked movies: \textbf{\underline{\textless disliked historical interactions\textgreater}}.
\newline Question: How would the user rank the candidate item list: \underline{\textless candidate item list\textgreater}?
\newline Hint 1: Another recommender model suggests \textbf{\underline{\textless ranking list\textgreater}}.
\newline Hint 2: These are corresponding entities and relationships for above model’s recommendation for more context information: \textbf{\textless KG sentence format information \textgreater}. 
\\ \hline

\end{tabular}
\label{tab1}
\end{table}

\subsubsection{User Sampling and Candidate Items Selection}\label{s3.4.1}
\zym{As \io{mentioned in the Introduction section}, the input token length of LLMs prevents directly inputting a large number of user interactions and a complete set of candidate items.}
\zym{Therefore, our aim is to obtain representative users through selective user sampling and to streamline the candidate item set to meet the input constraints of \io{an} LLM.}
\zym{We follow the same methods as~\cite{luo2023recranker} to perform user sampling and candidate item selection.}
\zym{Specifically, for user sampling, we first prioritise sampling from users with more interactions, which provide more reliable and consistent data.}
\zym{\io{Next,} we cluster user embeddings using the K-Means~\cite{hartigan1979algorithm} clustering algorithm and sample proportionally from each cluster, thereby 
preventing certain user groups from being over-concentrated
\z{obtaining representative users from different clusters.}
}
\zym{Finally, we perform probability decay\z{~\cite{luo2023improving}} for users selected 
\z{by clustering-based sampling}
to reduce their repeated sampling rate.}

\subsubsection{Instruction Prompt Construction}\label{s3.4.2}
Once we obtain the sampled users, we can combine them with the previously retrieved top-$Q$ KG triples set 
\zz{from the GraphRAG component}
to construct prompts.
Table~\ref{tab1} shows an illustrative listwise 
and two 
\zym{variants}
prompts \zym{({KERAG\_R}$_t$ and {KERAG\_R}$_s$)} that we designed. 
\zym{In general, we construct the knowledge-enhanced prompt with the following parts: the top-$k$ task description, the user's historical interactions, the task objective, rankings from another recommendation model, and the KG information.}
Specifically, 
\zym{as shown in Figure 1 (a),}
we retrieve the top-$Q$ triples with the higher similarity scores from the 
\zym{GraphRAG} for the  user's \zym{interacted}
items and candidate items, and integrate these triples into our 
\zym{prompt.}
Furthermore, in order to integrate 
\zy{collaborative}
signals from traditional recommendation models into 
\io{the} LLM for top-$k$ 
\zym{recommendation,}
we use LightGCN\z{~\cite{he2020lightgcn}},
\z{which effectively captures collaborative signals while maintaining computational efficiency,}
to predict 
users' 
\zm{preferences}
, and then convert these predictions into natural language descriptions 
\zym{and integrate them into prompts as a hint.}
\zym{As shown in Figure 1 (b)},
to determine the best representation format for integrating KG triples in the instruction prompt,
we \zm{design}
two \zym{types} knowledge-enhanced prompts with
\zym{relational knowledge between items and entities}
: the first \zym{type of} prompt incorporates the KG information obtained by 
\zym{knowledge-enhanced GraphRAG}
into the prompt in the format of triples, 
\zym{represented as}
"item text - relation text - entity text" 
\zym{(e.g.,}
Cameron - director\_film - The Terminator).
As for the second 
\zym{variant}
prompt, we incorporate
the KG information in the format of natural language sentences
\zym{(e.g.,}
Cameron is the director of The Terminator).
As such, we construct knowledge-enhanced prompts with relational information and use them for subsequent instruction tuning \zym{(see Figure 1 (c))}.
\subsection{\zym{Knowledge-Enhanced Instruction Tuning}}\label{s3.5}
As discussed in Section\zz{~\ref{s2.1}}, instruction tuning enables LLMs to better \io{capture} and follow 
task-specific instructions, enhancing their adaptability and performance across diverse downstream applications\zm{~\cite{zhang2023instruction,tang2024graphgpt}}. 
\zym{In this paper, we}
adopt a lightweight tuning method
\zym{- LoRA~\cite{hu2021lora} -}
to efficiently tune large language models (LLMs) 
\zym{with three parameters while maintaining effectiveness.}
\zym{The core premise of this lightweight tuning method is that current \io{LLMs} have an excessive number of parameters, with information concentrated in a lower intrinsic dimensionality~\cite{bao2023tallrec,zhao2023survey}.}
\zym{To efficiently adapt the 
\z{Llama-3.1-8B-Instruct~\cite{dubey2024llama}}
, we apply LoRA to introduce low-rank matrices to constrain \io{the} weight updates, significantly reducing computational overhead while preserving \io{the model's} effectiveness.}
In general, our objective is to fine-tune the 
\zym{Llama-3.1-8B-Instruct~\cite{dubey2024llama}}
\zym{using instruction tuning}
by minimising the cross entropy loss: 

\begin{equation}
\mathcal{L}=\min_\Theta\sum_{(x,y)\in\mathcal{D}_{train}}\sum_{t=1}^{|y|}-\log P_\Theta\left(y_t\mid x,y_{[1:t-1]}\right),
\end{equation}
where $\Theta$ represents the parameters of the LLM,
$\mathcal{D}_{train}$ is the training 
\zm{data,}
${|y|}$ is the length of the target sequence $y$, 
and $P_\Theta$ represents the probability of generating the $t$-th token $y_t$ 
in the target output $y$, given the input $x$ and the previous token $y$.
\z{We minimise this loss function to fine-tune the LLM parameters $\Theta$, enabling the LLM to capture user preferences expressed and structured relational knowledge in natural language.}
\z{This training effectively adapts the LLM to the recommendation system task.}
\zym{As such, the instruction-tuned LLM obtained through this process \io{allows to} incorporate domain-specific knowledge into \io{the} recommendation task, enabling the generation of recommendations that are more contextually aware and better aligned with \io{the} users' preferences, thereby improving the \io{recommendation} performance.}
\subsection{\zz{Efficiency Analysis}}\label{s3.6}
\zz{We analyse the efficiency of our KERAG\_R model by reporting the training time, inference time, and model size of its key components: GraphRAG and knowledge-enhanced instruction tuning.}
\zz{For knowledge-enhanced instruction tuning, we train Llama-3.1 (8B) on 1,000 instructions using a single A6000 GPU, which takes approximately 74 hours. 
The average inference time per instruction is around 6.2 seconds.} 
\zz{In comparison, RecRanker takes approximately 68 hours to train 1,000 instructions and about 4.2 seconds per instruction for inference on the same hardware.}
\zz{This demonstrates that our model maintains comparable compute usage.}
\zz{Our backbone LLM (Llama-3.1) contains 8 billion parameters, while the additional GAT model used in GraphRAG introduces only 
\zz{0.68 million trainable}
parameters, making its contribution to the total model size marginal in comparison. 
Thus, the overall computational cost remains dominated by the LLM.} 
\zz{In summary, KERAG\_R maintains comparable training and inference efficiency to RecRanker}
\zz{while 
incorporating a lightweight GAT-based retrieval module 
\zz{that only requires marginal additional parameters.}
}

\section{Experiments}\label{s4}
We conduct experiments on three datasets to evaluate the performance of our 
\zy{KERAG\_R}
model.
We compare with several different groups of models, addressing 
\z{three}
research questions:

\noindent \textbf{RQ1}: How does 
\zy{KERAG\_R}
perform compared to existing 
\z{state-of-the-art}
recommendation models?

\noindent \textbf{RQ2}: 
\zm{How do the main components (LLM, GraphRAG, KG triple selection \zym{and knowledge-enhanced instruction tuning}) of KERAG\_R affect the recommendation performance?}

\noindent \textbf{RQ3}: 
How \z{does} 
\z{the selection of}
different numbers of KG triples affect KERAG\_R's performance? \\

\subsection{Experimental Settings}\label{s4.1}
\subsubsection{Datasets}\label{s4.4.1}
\begin{table}[tb]
\centering
\caption{Statistics of the used datasets.}
\resizebox{0.8\linewidth}{!}{ 
\begin{tabular}{cccc}
     \hline
     & ML-1M & ML-10M & Amazon-book\\
     \hline \hline 
     Users &6,040  &71,567  &28,104 \\
     Items &3,952  &10,681  &24,903 \\
     Interactions &1,000,209  &10,000,054  &582,321 \\
     \hline
     Interaction density(\%) &0.0419  &0.0131  &0.0008 \\     
     \hline
     Entities &31,380  &32,754  &23,780 \\
     Relations &31  &31  &10 \\
     Triples &70,444  &133,245  &102,149 \\
     \hline
\end{tabular}
}
\label{tab2}
\vspace{-4mm}
\end{table}
Following~\cite{bao2023tallrec,luo2023recranker}, we conduct experiments to evaluate our \zy{KERAG\_R} 
\zm{model}
on three 
public \io{and} 
\zm{widely-used} 
datasets, namely MovieLens-1M (ML-1M)\footnote{\small{https://grouplens.org/datasets/movielens/}}, 
\zy{MovieLens-10m}
(ML-10M)\footnote{\small{https://grouplens.org/datasets/movielens/}} and Amazon-book\footnote{\small{http://jmcauley.ucsd.edu/data/amazon}}. 
The MovieLens dataset is 
used as a standard benchmark for movie  
\zym{recommendation}, \io{and} includes \zy{users'} ratings of movies.
We use two 
\zy{datasets from}
MovieLens
\zy{~\cite{harper2015movielens}}: MovieLens-1M (ML-1M), which contains about 1,000,000 user ratings of movie items, and ML-10M (ML-10M), which is expanded to more than 10,000,000 ratings.
As for the Amazon-book dataset, it contains user ratings from 1 to 5 and is often used for book recommendation tasks.
\io{For} these three datasets, we follow \io{the} 
\zy{same}
settings \io{as}~\cite{luo2023recranker}, where we remove \io{users and items} with less than 10 interactions in the historical interaction information \io{in order} to improve the quality of the dataset. 
The 
statistics of the datasets\zm{, which are also widely used for KG-based recommendations}, are shown in Table~\ref{tab2}.
For these three datasets, we map 
\zm{all items}
to the Freebase entities via title matching to obtain \io{the} corresponding knowledge graphs~\cite{bollacker2008freebase,zhao2019kb4rec}. 

\subsubsection{Data Preprocessing}\label{s4.1.2}
We evaluate 
\z{all models}
using a leave-one-out strategy~\cite{luo2023recranker,hu2024enhancing}, partitioning the dataset consistently with previous studies.
\zm{This strategy typically splits the dataset based on timestamps: for each user, the most recent interaction is considered as the test instance, the second most recent interaction is used for validation, and all prior interactions form the training set.}
\zm{In this work, following RecRanker~\cite{luo2023recranker}, we sample 1,000 user interactions as instructions from each dataset to fine-tune the LLM.}
During \io{the} instruction tuning process,
\z{we also follow~\cite{luo2023recranker} to}
construct a candidate list by randomly selecting three highest-rated items, two second-highest-rated items, and five unrated items,
\z{and then we}
select the top five highest-rated items as ground truth.
\zym{The preprocessing scripts we use are \io{given} in the provided anonymous repository \z{linked in the abstract.}}

\pageenlarge{1}
\subsubsection{\zym{Evaluation Protocol and Implementation Details}}\label{s4.1.3}
\zym{Following \io{previous} studies~\cite{bao2023tallrec,luo2023recranker},  we use two representative metrics, \io{namely} Hit Ratio (HR) and Normalized Discounted Cumulative Gain (NDCG), to evaluate the effectiveness of our KERAG\_R model \z{in comparison to the baselines}.}
\zym{Existing LLM-based models~\cite{zhang2023recommendation,luo2023recranker} typically set $k$ 
\z{, the number of candidate items used in top-$k$ evaluation metrics,}
to 3 and 5 due to the input token length limitations of LLMs. 
Following this approach, we use the same settings in our experiments.}
We choose Llama-3.1-8B-Instruct~\cite{dubey2024llama} 
for experiments because of its strong performance among open source models. 
\io{Indeed, we} also \io{experimented with} several open-source LLMs 
such as Llama-2 (7B) and Llama-2 (13B)~\cite{touvron2023llama}, 
\io{but Llama-3.1-8B-Instruct}
was consistently \io{the most effective}.
\zm{For the GAT pre-training, we set the input embedding dimension to 16, batch size and chunk size to 10 and 50 respectively.}
During the training phase of Llama-3.1(8B), we uniformly adopt a learning rate of 2e-5 and a context length of 2048.
We train in epochs and implement a cosine scheduler, integrating a preliminary warm-up phase of 50 steps.
We use LoRA~\cite{hu2021lora} to efficiently train these models.
\zym{For the rank size $r$ and scaling factor alpha of LoRA, we vary this value within \{8, 16, 32, 64\}.}
In the inference phase, we \z{follow~\cite{luo2023recranker}} to use the vLLM~\cite{kwon2023efficient} framework, setting the temperature parameter to 0.1, top-$k$ sampling to 40, and nucleus sampling to 0.1. 
\zy{We optimise our KERAG\_R model and baselines using the Adam optimiser\cite{kingma2014adam}, and initialise \io{the} embeddings with Xavier initialisation~\cite{glorot2010understanding}.}

\pageenlarge{1}
\subsubsection{Baselines:}\label{s4.1.4}
We compare our KERAG\_R model
with four different groups of \io{existing} baselines: \\
(1) Collaborative filtering recommendation models: 
\zz{\textbf{MF}~\cite{koren2009matrix}}
\zym{
is the first model to}
decompose the user-item interaction matrix into two lower-dimensional matrices to capture hidden features that drive user preferences and item characteristics; 
\zz{\textbf{LightGCN}~\cite{he2020lightgcn}}
\zym{simplifies the convolution operations for \io{the} information propagation between users and items by removing nonlinear activations and transformation matrices, resulting in \io{an} improved effectiveness;}
\zz{\textbf{MixGCF}~\cite{huang2021mixgcf}}
\zym{generates synthetic negative samples by aggregating embeddings from different layers in the neighbourhood of the original negative samples, effectively extracting negative signals from implicit feedback.} \\
\noindent (2) Sequential Recommendation Models:
\zz{\textbf{SASRec}~\cite{kang2018self}}
leverages a masked multi-head attention mechanism to model \io{the users'} sequential interactions;
\zz{\textbf{BERT4Rec}~\cite{sun2019bert4rec}}
\zym{
BERT4Rec is the first to use deep bidirectional self-attention to capture the sequential information of user interactions;}
\zz{\textbf{CL4SRec}~\cite{xie2022contrastive}}
obtains self-supervisory signals from the original user behaviour sequences. Different from SASRec and BERT4Rec, CL4SRec uses three data augmentation methods to enhance \io{the} users' representations.\looseness -1 \\
\noindent (3) Knowledge Graph Recommendation Models:
\zz{\textbf{CKE}~\cite{zhang2016collaborative}}
uses a heterogeneous network embedding approach, extracting structural representations of items by considering the heterogeneity of nodes and relations, thereby improving the learning of user preferences;
\zz{\textbf{KGAT}~\cite{wang2019kgat}}
distinguishes itself from CKE by emphasising the importance of neighbouring nodes in the knowledge graph through an attention mechanism and enriches item representation learning by aggregating information from adjacent nodes \io{to} the item.\looseness -1 \\
\noindent (4) LLM\zym{-based} Recommendation Model:
\zz{\textbf{P5}~\cite{geng2022recommendation}}
\zz{
is a T5-based~\cite{raffel2020exploring} encoder–decoder generative model that unifies various recommendation tasks into a single LLM framework.}
\zz{We adopt OpenP5~\cite{xu2024openp5}, the open-source implementation of P5, 
and use the variant 
\zz{designed for the top-$k$ recommendation task}
as our baseline;} 
\zz{\textbf{RecRanker}~\cite{luo2023recranker}}
\zym{
\io{uses} adaptive user sampling to
\zz{construct}
prompts and leverages an instruction-tuned LLM to perform recommendations by combining three types of ranking tasks in top-$k$ recommendation.}
\begin{table*}[tb]
\centering
\caption{Performance comparison results between two variants of KERAG\_R and other baselines on three datasets. The best result is bolded and the second best result is underlined. The superscript $^*$ indicates that the results are significantly different from the {KERAG\_R}$_t$ results using the Holm-Bonferroni corrected paired t-test with p-value<0.05.} 
\resizebox{\textwidth}{!}{
\begin{tabular}{ccccccccccccc}
\hline
\textbf{Dataset} & \multicolumn{4}{c}{ML-1M} & \multicolumn{4}{c}{ML-10M} & \multicolumn{4}{c}{Amazon-Book} \\
\cmidrule(lr){1-1} \cmidrule(lr){2-5} \cmidrule(lr){6-9} \cmidrule(lr){10-13} 
Methods & HR@3 & NDCG@3  & HR@5 & NDCG@5 & HR@3 & NDCG@3  & HR@5 & NDCG@5 & HR@3 & NDCG@3  & HR@5 & NDCG@5\\
\hline
MF & {0.0224}$^*$ & {0.0163}$^*$ & {0.0363}$^*$ & {0.0220}$^*$ & {0.0273}$^*$ & {0.0199}$^*$ & {0.0420}$^*$ & {0.0259}$^*$ & {0.0189}$^*$ & {0.0136}$^*$ & {0.0296}$^*$ & {0.0180}$^*$\\
LightGCN & {0.0250}$^*$ & {0.0182}$^*$ & {0.0413}$^*$ & {0.0251}$^*$ & {0.0285}$^*$ & {0.0205}$^*$ & {0.0459}$^*$ & {0.0281}$^*$ & {0.0222}$^*$ & {0.0161}$^*$ & {0.0330}$^*$ & {0.0205}$^*$\\
MixGCF & {0.0184}$^*$ & {0.0130}$^*$ & {0.0274}$^*$ & {0.0166}$^*$ & {0.0221}$^*$ & {0.0155}$^*$ & {0.0413}$^*$ & {0.0193}$^*$ & {0.0151}$^*$ & {0.0104}$^*$ & {0.0208}$^*$ & {0.0149}$^*$\\
SASRec & {0.0139}$^*$ & {0.0095}$^*$ & {0.0219}$^*$ & {0.0127}$^*$ & {0.0233}$^*$ & {0.0150}$^*$ & {0.0386}$^*$ & {0.0213}$^*$ & {0.0112}$^*$ & {0.0064}$^*$ & {0.0191}$^*$ & {0.0096}$^*$\\
BERT4Rec & {0.0096}$^*$ & {0.0065}$^*$ & {0.0174}$^*$ & {0.0098}$^*$ & {0.0104}$^*$ & {0.0072}$^*$ & {0.0172}$^*$ & {0.0100}$^*$ & {0.0101}$^*$ & {0.0065}$^*$ & {0.0192}$^*$ & {0.0103}$^*$\\
CL4SRec & {0.0129}$^*$ & {0.0089}$^*$ & {0.0192}$^*$ & {0.0115}$^*$ & {0.0229}$^*$ & {0.0149}$^*$ & {0.0382}$^*$ & {0.0211}$^*$ & {0.0093}$^*$ & {0.0054}$^*$ & {0.0175}$^*$ & {0.0087}$^*$\\
CKE & {0.0267}$^*$ & {0.0194}$^*$ & {0.0378}$^*$ & {0.0239}$^*$ & {0.0288}$^*$ & {0.0209}$^*$ & {0.0455}$^*$ & {0.0277}$^*$ & {0.0231}$^*$ & {0.0169}$^*$ & {0.0351}$^*$ & {0.0218}$^*$\\
KGAT & {0.0265}$^*$ & {0.0194}$^*$ & {0.0404}$^*$ & {0.0250}$^*$& {0.0289}$^*$ & {0.0208}$^*$ & {0.0460}$^*$ & {0.0281}$^*$ & {0.0240}$^*$ & {0.0173}$^*$ & {0.0362}$^*$ & {0.0223}$^*$\\
P5 & {0.0194}$^*$ & {0.0145}$^*$ & {0.0303}$^*$ & {0.0190}$^*$& {0.0273}$^*$ & {0.0190}$^*$ & {0.0470}$^*$ & {0.0270}$^*$ & {0.0020}$^*$ & {0.0038}$^*$ & {0.0015}$^*$ & {0.0023}$^*$\\
RecRanker & \underline{0.0270}$^*$ & \underline{0.0200}$^*$ & \underline{0.0430}$^*$ & \underline{0.0262}$^*$ & \underline{0.0304}$^*$ & \underline{0.0219}$^*$ & \underline{0.0473}$^*$ & \underline{0.0289}$^*$ & \underline{0.0259}$^*$ & \underline{0.0188}$^*$ & \underline{0.0401}$^*$ & \underline{0.0246}$^*$\\
\hline
{KERAG\_R}$_s$ & {0.0287}$^*$ & {0.0210}$^*$ & {0.0469}$^*$ & {0.0284}$^*$ & {0.0324}$^*$ & {0.0240}$^*$ & {0.0492}$^*$ & {0.0309}$^*$ & {0.0283}$^*$ & {0.0210}$^*$ & {0.0430}$^*$ & {0.0269}$^*$\\
{KERAG\_R}$_t$ & \textbf{0.0293} & \textbf{0.0219} & \textbf{0.0478} & \textbf{0.0293} & \textbf{0.0336} & \textbf{0.0249} & \textbf{0.0514} & \textbf{0.0321} & \textbf{0.0290} & \textbf{0.0216} & \textbf{0.0438} & \textbf{0.0277}\\
\hline
Improve\% & {+8.52\%} & {+9.50\%} & {+11.16\%} & {+11.83\%} & {+10.53\%} & {+13.70\%} & {+8.67\%} & {+11.07\%} & {+11.97\%} & {+14.89\%} & {+7.23\%} & {+9.35\%}\\
\hline
\end{tabular}
}
\label{tab3}
\end{table*}

\zz{We omit the comparison with other LLM-based recommendation models such as TALLRec~\cite{bao2023tallrec}, InstructRec~\cite{zhang2023recommendation}, and LLMRec~\cite{wei2024llmrec} due to significant differences in task settings, availability, or the lack of open-source implementations.}
\zz{Specifically, TALLRec models the sequence of user interactions using the Llama-7B as its backbone in a sequential recommendation setting, which differs from our top-$k$ recommendation task.}
\zz{InstructRec relies on a closed-source GPT model that does not support instruction tuning, making it incompatible with our experimental setup that requires full access for model adaptation.}
\zz{LLMRec lacks publicly available implementation details and key preprocessing scripts, preventing full replication of its pipeline.}

\subsection{Performance Comparison (RQ1)}\label{s4.2}

Table~\ref{tab3} reports the results of comparing our two KERAG\_R variants ({KERAG\_R}$_t$, {KERAG\_R}$_s$) with all the baselines described in Section\zz{~\ref{s4.1.4}}. We evaluate {KERAG\_R}$_t$ and {KERAG\_R}$_s$ in comparison to four distinct recommendation approaches: collaborative filtering models (MF, LightGCN, MixGCF), sequential models (SASRec, BERT4Rec, CL4SRec), KG models (CKE, KGAT)
\zz{and LLM-based models (P5, RecRanker).}
\io{Recall that} {KERAG\_R}$_t$ incorporates relational information into the instruction prompt using a triple representation, while {KERAG\_R}$_s$ uses a sentence representation. \io{In Table~\ref{tab3}}, 
the best and second-best performances are indicated in bold and underlined, respectively.
From the results in Table~\ref{tab3}, we observe the following:
\pageenlarge{1}
\begin{itemize}[leftmargin=*, noitemsep, topsep=0pt, partopsep=0pt]
    \item For all three datasets, we \io{observe} that both the {KERAG\_R}$_t$ and {KERAG\_R}$_s$ variants significantly outperform all groups of baselines on both \io{used} metrics as confirmed by a paired t-test with the Holm-Bonferroni correction. The improvements can be attributed to the integration of the GraphRAG component and the knowledge-enhanced instruction tuning method, which incorporate domain-specific knowledge from the KG into LLM prompts.
    Notably, our KERAG\_R variants also outperform the strongest LLM-based baseline, RecRanker, on all used datasets by a large margin.
    This result indicates that incorporating relational knowledge (i.e., item-entity triples) during the instruction tuning stage can improve the effectiveness of the LLM-based recommendation.
    \pageenlarge{2}
    \item By comparing the {KERAG\_R}$_t$ and {KERAG\_R}$_s$ variants, we observe from Table~\ref{tab3} that {KERAG\_R}$_t$ outperforms {KERAG\_R}$_s$ in all instances.
    These results indicate that presenting relational information in the format of triples is more effective than \io{using the} natural language sentences. 
    \io{The} advantage of triples lies in their conciseness and the reduced presence of non-relevant information, 
    which enriches the context provided for the users' interacted items, hence allowing the used LLM to more effectively estimate the user preferences.
    A recent study~\cite{fang2024trace} also \io{made a similar observation on a different task, namely} that KG triples provide more accurate relational knowledge than sentences in question-answering tasks.
    \item Table~\ref{tab3} also shows that the KG models (CKE, KGAT) and the LLM model (RecRanker) are generally more effective than the collaborative filtering models (MF, LightGCN, MixGCF) and the sequential models (SASRec, BERT4Rec, CL4SRec). This result indicates that it is beneficial to introduce additional knowledge (i.e., a KG) to enrich the user/item representations in the top-$k$ recommendation task. 
    
\end{itemize}
In answer to RQ1, we conclude that both our KERAG\_R variants successfully leverage GraphRAG and knowledge-enhanced instruction tuning to enhance the recommendation performance and outperform all the existing strong baselines.
\vspace{-3mm}

\subsection{Ablation Study (RQ2)}\label{s4.3}
\pageenlarge{3}
\begin{table*}
\centering
\caption{Results of the ablation study. The $^*$ indicates that the result is significantly different from {KERAG\_R}$_t$ using the Holm-Bonferroni corrected paired t-test with p-value<0.05.} 
\resizebox{\textwidth}{!}{
\begin{tabular}{ccccccccccccc}
\hline
\textbf{Dataset} & \multicolumn{4}{c}{ML-1M} & \multicolumn{4}{c}{ML-10M} & \multicolumn{4}{c}{Amazon-Book} \\
\cmidrule(lr){1-1} \cmidrule(lr){2-5} \cmidrule(lr){6-9} \cmidrule(lr){10-13} 
Methods & HR@3 & NDCG@3  & HR@5 & NDCG@5 & HR@3 & NDCG@3  & HR@5 & NDCG@5 & HR@3 & NDCG@3  & HR@5 & NDCG@5\\
\hline
$w/o-Llama3 $ & {0.0238}$^*$ & {0.0171}$^*$ & {0.0421}$^*$ & {0.0245}$^*$& {0.0271}$^*$ & {0.0194}$^*$ & {0.0434}$^*$ & {0.0261}$^*$ & {0.0242}$^*$ & {0.0175}$^*$ & {0.0393}$^*$ & {0.0240}$^*$\\
$w/o-graphrag$ & {0.0277}$^*$ & {0.0203}$^*$ & {0.0442}$^*$ & {0.0270}$^*$ & {0.0307}$^*$ & {0.0221}$^*$ & {0.0475}$^*$ & {0.0290}$^*$ & {0.0277}$^*$ & {0.0203}$^*$ & {0.0424}$^*$ & {0.0263}$^*$\\
$w/o-instruction$ & {0.0282}$^*$ & {0.0205}$^*$ & {0.0436}$^*$ & {0.0267}$^*$ & {0.0309}$^*$ & {0.0225}$^*$ & {0.0476}$^*$ & {0.0294}$^*$ & {0.0280}$^*$ & {0.0210}$^*$ & {0.0429}$^*$ & {0.0266}$^*$\\
$w/o-selection$ & \underline{0.0286}$^*$ & \underline{0.0207}$^*$ & \underline{0.0455}$^*$ & \underline{0.0280}$^*$ & \underline{0.0320}$^*$ & \underline{0.0232}$^*$ & \underline{0.0491}$^*$ & \underline{0.0306}$^*$ & \underline{0.0286}$^*$ & \underline{0.0214}$^*$ & \underline{0.0430}$^*$ & \underline{0.0269}$^*$\\
\hline
{KERAG\_R}$_t$ & \textbf{0.0293} & \textbf{0.0219} & \textbf{0.0478} & \textbf{0.0293}  & \textbf{0.0336} & \textbf{0.0249} & \textbf{0.0514} & \textbf{0.0321} & \textbf{0.0290} & \textbf{0.0216} & \textbf{0.0438} & \textbf{0.0277}\\
\hline
\end{tabular}
}
\label{tab4}
\end{table*}
\begin{figure*}
\centering
\includegraphics[width=0.8\linewidth]{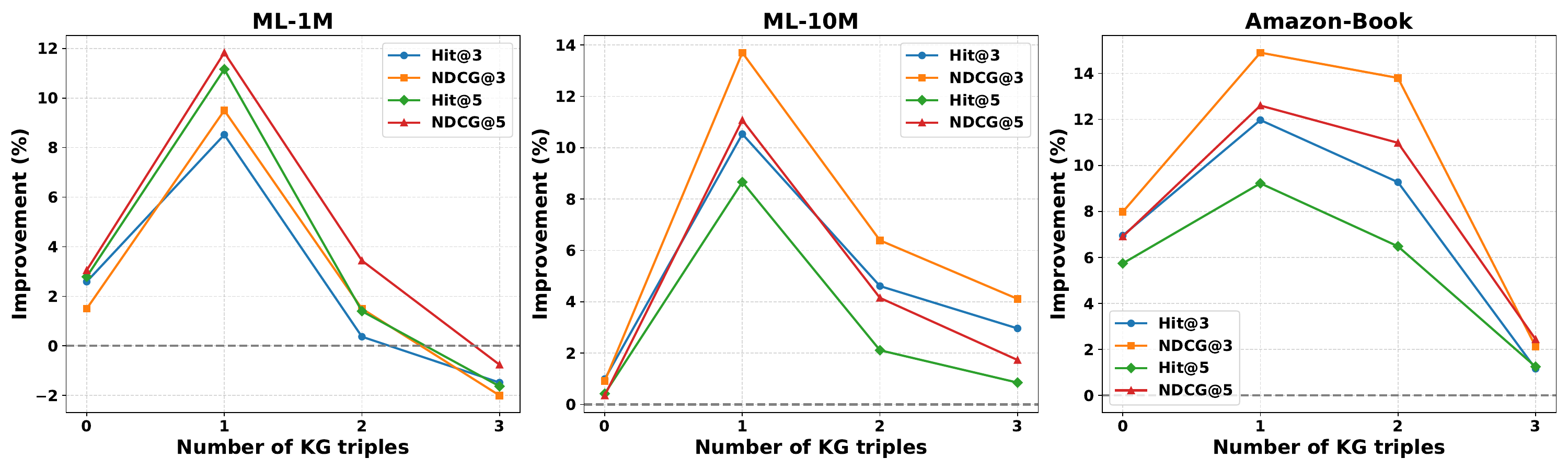}
\caption{Performance of our KERAG\_R model with respect to different numbers of triples.} 
\label{fig2}
\end{figure*}
In this section, we ablate each of the key components of our proposed KERAG\_R model by introducing four variants, namely (i) $w/o-graphrag$ (ii) $w/o-instruction$ (iii) $w/o-selection$ and (iv) $w/o-Llama3$. In particular,  
$w/o-graphrag$ is a variant that removes the \zym{GraphRAG} component; 
\zym{$w/o-instruction$ is a variant that removes the knowledge-enhanced instruction tuning component;}
$w/o-selection$ is a variant that replaces the use of the 
\zym{knowledge-enhanced}
triple selection method with a random selection of triples from the KG;
$w/o-Llama3$ is a variant that replaces the used LLM, Llama-3.1-8B-Instruct, with Llama-2 (7B)~\cite{touvron2023llama}.
We compare these variants to our {KERAG\_R}$_t$ variant due to its promising effectiveness, as observed in Table~\ref{tab3}. \io{Note that} the same conclusions \io{are drawn with the} {KERAG\_R}$_s$ variant.

\begin{itemize}[leftmargin=*, noitemsep, topsep=0pt, partopsep=0pt]
    \item To explore the impact of GraphRAG in our KERAG\_R model, we remove the GraphRAG component and maintain the original instruction tuning for the top-$k$ recommendation task. We observe from Table~\ref{tab4} that the $w/o-graphrag$ variant underperforms {KERAG\_R}$_t$ on all three datasets. This result indicates the importance of incorporating domain-specific knowledge in the instruction prompt to effectively guide the used LLM in estimating the users' potential interactions.
    \item As discussed in Section\zz{~\ref{s3.3.2}}, we aim to investigate the effectiveness of our triple selection method for GraphRAG. We substitute the top-Q selection method with a random selection for each user's interacted items.
    Table~\ref{tab4} shows that {KERAG\_R}$_t$ outperforms $w/o-selection$ across all used datasets by a large margin. This superior performance demonstrates the suitability of selecting the most relevant triples for each target user's interacted items, thereby enhancing the LLM's reasoning ability in the top-$k$ recommendation scenario.
    \item To investigate the effectiveness of our knowledge-enhanced instruction tuning component, we remove this component and use Llama-3 in a zero-shot configuration for the top-$k$ recommendation task. As shown in Table~\ref{tab4}, \io{we} observe that the $w/o-instruction$ variant underperforms {KERAG\_R}$_t$.
    This result indicates the importance of instruction tuning in guiding the LLM using additional knowledge, thereby enhancing the LLM's adaptability to the recommendation scenario at hand.
    \item To investigate the usefulness of the used LLM, Llama-3, we conduct a comparative analysis by replacing the Llama-3 model with Llama-2 as the used LLM.
    From Table~\ref{tab4}, we observe that {KERAG\_R}$_t$ significantly outperforms $w/o-Llama3$ across all datasets. This finding confirms the effectiveness of an instruction-tuned Llama-3 in facilitating reasoning within the top-$k$ recommendation task.
\end{itemize}
Overall, in response to RQ2, we conclude that our {KERAG\_R} model successfully leverages each of its components to provide an effective LLM-based approach for an enhanced top-$k$ recommendation.

\pageenlarge{1}
\subsection{Impact of the Number of Retrieved KG Triples (RQ3)}\label{s4.4}
We now examine the impact of the number of retrieved KG triples on the performance of our KERAG\_R model within the proposed GraphRAG component. For brevity, we report the results of the {KERAG\_R}$_t$ variant since we observe similar conclusions on the {KERAG\_R}$_s$ variant.
As mentioned in Section 3.3.2, we use a GAT model to retrieve the top-Q triples for each item. We conduct this analysis by varying the number of triples retrieved for each of the user's interacted items in a range of \{0, 1, 2, 3\}. 
In our initial experiments, we noted that retrieving more than four triples results in out-of-memory errors on a single GPU due to longer input sequences.
Figure~\ref{fig2} \io{shows} that our KERAG\_R model consistently achieves its peak performance when retrieving one triple across all three datasets. 
In particular, we observe performance decreases when adding additional triples for each item.
These results suggest that a single triple provides sufficient relational knowledge to effectively enhance the LLM's reasoning capabilities for the top-$k$ recommendation task. In contrast, integrating more triples in GraphRAG might lead to noise and redundancy that could detract the LLM's decision-making process.

In response to RQ3, we conclude that a single triple per user-interacted item is sufficiently effective.
Conversely, additional triples can hinder the LLM's reasoning process in top-$k$ recommendation. \looseness -1
\pageenlarge{1}
\vspace{-2mm}

\section{Conclusions}\label{s5}
\looseness -1 We proposed a novel recommendation model called Knowledge-Enhanced
Retrieval-Augmented Generation for Recommendation (KERAG\_R)
to address the absence of domain-specific knowledge in the existing LLM-based recommendation models for top-$k$ recommendation.
Specifically, KERAG\_R included a novel GraphRAG component and a knowledge-enhanced instruction tuning method to integrate external knowledge from a knowledge graph (KG) into the LLM (i.e., Llama-3) prompts, thereby enhancing the LLM’s reasoning in the top-$k$ recommendation task.
Our extensive experiments on three datasets showed that our KERAG\_R model significantly outperformed 
\zz{ten}
strong recommendation baselines, including the existing state-of-the-art LLM-based model, RecRanker.
In addition, our ablation study showed the positive impact of each component
of our KERAG\_R model on the recommendation performance. We also conducted a study to determine the best number of triples for each user's interacted item within the used LLM's input token length. In particular, we found that retrieving the most relevant KG
information is more effective than using additional KG triples, and
that relational KG triple representations outperform natural KG sentence representations in the prompts.


\clearpage
\bibliographystyle{ACM-Reference-Format}
\bibliography{reference}
\end{document}